# An Attempt at Analyzing the Information Nature of Money[1]

Haibo Chen[2]

**Abstract:**

Money was invented to address the difficulty in the double coincidence of wants between the supply and demand when people exchanged their goods and services. There are two information states in society: one is the initial state that people have goods and services due to division of labor; the other is the final state that people have different goods and services with the initial state due to exchange of goods and services between them. The process is that the initial state is changed to the final state with the help of money. Because the direct exchanges of goods and services are difficult to achieve the double coincidence of wants in time and space, it can be achieved with the help of money which is as a medium and bridge. In this paper the changing process of the state information is analyzed through the matrix representation of money, and then the nature of money with a kind of information of reliable ledger is more apparently shown. This paper also analyzes the common characteristics of physical money, electronic money and digital currency, that is, reliable ledger and explores the future trend of money development from the perspective of history and security technology of money.

**Keywords:** money, physical money, electronic money, digital currency, centralized physical ledger, distributed physical ledger, electronic ledger, distributed ledger.

## 1 Introduction

At present, with the rapid development of electronic computers and communication technologies, non-cash payments such as electronic payment and mobile payment are becoming more and more popular in people's lives. In recent years, the developments of electronic payments in Nordic countries has been particularly rapid, and even the slogan of a cashless society has been proposed. However, the demand for global cash has increased year by year, and cash still has a strong vitality. In such an era of diversified money forms and multiple possibilities for future money development, the nature of money needs further analysis.

From the perspective of currency history and currency anti-counterfeiting technology, this paper analyzes the common characteristics and respective advantages of physical money, electronic money, and digital currency, and explores the future development trend of money. The full paper also includes the following sections: Section II, the definitions of physical money, electronic money, and digital currency. Section III starts from the fundamental problem, proposes the concept of money matrix, and intuitively reveals that money essence is information, which is an important feature of reliable ledger information; Section IV analysis from the history of money, the money may originate from the centralized physical ledgers; In section V, by studying the history of money development and the anti-counterfeiting technology of physical money, physical

---

money can be viewed as a distributed physical ledger with a large enough space size; Section VI, through the comparison of electronic and physical ledgers, electronic money can be seen as the electronic form of physical ledgers; In section VII, by comparing digital currency ledger and electronic money ledger, digital currency can be thought of as a kind of electronic money with multiple parties participating in bookkeeping; finally, through the comparison of various forms of money, we believe that the future trend of currency development is the coexistence and integration of multiple forms of ledgers.

2 **The definition of money**

Given that there are many kinds of money classifications, and in many cases the same name, the content of definition is not the same. This paper, generally, defines the physical money, electronic money, and digital currency as follows.

Before definition, it is very likely that money cannot be looked at in isolation. Taking money as a general equivalent, a string of deposit data may not be suitable. Because money is inseparable from the owner or holder and is provided for use by a group of people of a certain range, the money needs to be seen in the systematical and integrated view. For example: gold can't be called money if it's not mined; a banknote that is not issued from the central bank at the banknote printing factory can only be called a product, not be called money. The money may be more properly understood from the perspective of the ledger, and its three attributes can be understood as follows: the store of value means the money ledger is accurate, long-term, and cannot be maliciously tampered with; the medium of exchange means the money ledger information updates according to the real economic activities; the unit of value means that people reach a unit of value measurement through consensus, which can be the weight of the grain, or the weight of the gold, or even a virtual unit of value, such as "yuan", for value comparison and statistics.

Physical money is the tangible money, a distributed physical ledger that people can touch and see, such as commodity money, gold, silver, banknotes, coins, and so on. Electronic money is the intangible money, a ledger recorded electronically on an electronic device etc., especially hard disk with ferromagnetic materials. Under the protection of the encryption mechanism, the user's electronic identity authentication information is transmitted through the communication technology, and the user accesses the ledger so that the ledger information is not tampered with by malicious persons, and the update of its information truly reflects the economic activity status. Digital currency is the intangible money based on blockchain technology, or distributed ledger technology (Shrier et al (2016)). Digital currency can be understood as one kind of electronic money, which will be explained later.

3 **Nature of money is a kind of information**

According to Wikipedia's definition of money, money is any verifiable record that is widely accepted in specific countries or socio-economic entities, paid for goods and services, and repaid for debt. Money is a kind of information, in essence, a kind of reliable ledger information with many records.

Money is a great invention in human history. As Adam Smith states in "An Inquiry into the Nature and Causes of the Wealth of Nations" (Smith (2007)), the greatest improvement in the productive powers of labor seem to have been the effects of the division of labor. After the division of labor, people own different goods and services, which can be regarded as an information state of human society. In order to pursue a happy life, people need to change the state of such information and reach another state of information by sharing the goods and services. When the direct exchange of goods is difficult to meet the double coincidence of wants in time and space, in order to change the state of human information, the money was invented. Money is essentially information, as a medium and bridge to achieve the multi-party and indirect exchange and sharing of people's goods and services, changing the state of information of human society. For example, Alice produced 30 pieces of clothes, Bob produced 300 pounds of grains and Charlie built 3 houses.

| Alice | 30 clothes |
|---|---|
| Bob | 300 grains |
| Charlie | 3 houses |

Table 1, after division of labor, the initial state of goods

Alice, respectively, gives Bob and Charlie 10 clothes; Bob, respectively, gives Alice and Charlie 100 pounds of grains; Charlie, respectively, gives Alice and Bob 1 houses.

| Alice | 10 clothes, 100 grains, 1 house |
|---|---|
| Bob | 10 clothes, 100 grains, 1 house |
| Charlie | 10 clothes, 100 grains, 1 house |

Table 2, the final state of goods after exchange with the help of money as a medium and bridge,

Through the exchange, Alice, Bob and Charlie shared their goods. The above transaction process can be expressed by a matrix of general relations:

$$A+M = C$$

Where A is the initial state matrix, M is the money matrix, and C is the final state matrix. For the above transaction, the specific matrix relationship is expressed as follows:

$$\begin{bmatrix} 3 & 0 & 0 \\ 0 & 3 & 0 \\ 0 & 0 & 3 \end{bmatrix} + \begin{bmatrix} -2 & 1 & 1 \\ 1 & -2 & 1 \\ 1 & 1 & -2 \end{bmatrix} = \begin{bmatrix} 1 & 1 & 1 \\ 1 & 1 & 1 \\ 1 & 1 & 1 \end{bmatrix}$$

For convenience, the basic unit is set to 10 pieces of clothes, 100 pounds of grains, one house, where the rows of the matrix from small to large in turn said Alice, Bob and Charlie, the matrix columns from small to large in order of clothes, grain and house, then $\begin{bmatrix} 3 & 0 & 0 \\ 0 & 3 & 0 \\ 0 & 0 & 3 \end{bmatrix}$ is the initial state, in which $A_{11} = 3$, represents Alice has 30 clothes; $A_{22} = 3$, represents Bob has 300 pounds of grains; $A_{33} = 3$ said Charlie has three houses. $C=\begin{bmatrix} 1 & 1 & 1 \\ 1 & 1 & 1 \\ 1 & 1 & 1 \end{bmatrix}$ is the final state, that is, Alice, Bob and Charlie have 10 pieces of clothes, 100 pounds of grains and a house.

$M=\begin{bmatrix} -2 & 1 & 1 \\ 1 & -2 & 1 \\ 1 & 1 & -2 \end{bmatrix}$ denotes a money matrix, which consists of various transaction records, for example, $M_{21}=1$ can be represented as Alice sells to Bob 10 clothes. The matrix M contains various transaction records, which realizes the change of the information state from the initial state A to the final state B. It is a medium and bridge for people to share goods and services and represents money. In terms of mathematics, the nature of the matrix M, initial information state A and final information state B is also the same, which also reveals that money is the same as status information, essence is information, and is the ledger information of many transaction records.

This example is a bit special, and the three parties can reach the double coincidence of wants between supply and demand. In reality, this coincidence is not so easy to happen. However, from the point of view of the entire economic activity, the supply and demand side is still coupled. Therefore, many transactions between supply and demand are realized by a serial of the single transaction record. So the general relational expression can be:

$$\begin{bmatrix} A_{11} & \cdots & A_{1n} \\ \vdots & \ddots & \vdots \\ A_{n1} & \cdots & A_{nn} \end{bmatrix} + \begin{bmatrix} M_{11} & \cdots & M_{1n} \\ \vdots & \ddots & \vdots \\ M_{n1} & \cdots & M_{nn} \end{bmatrix}$$

$$= \begin{bmatrix} A_{11} & \cdots & A_{1n} \\ \vdots & \ddots & \vdots \\ A_{n1} & \cdots & A_{nn} \end{bmatrix} + \begin{bmatrix} M^1_{11} & \cdots & M^1_{1n} \\ \vdots & \ddots & \vdots \\ M^1_{n1} & \cdots & M^1_{nn} \end{bmatrix} \cdots$$

$$+ \cdots \begin{bmatrix} M^n_{11} & \cdots & M^n_{1n} \\ \vdots & \ddots & \vdots \\ M^n_{n1} & \cdots & M^n_{nn} \end{bmatrix} = \begin{bmatrix} C_{11} & \cdots & C_{1n} \\ \vdots & \ddots & \vdots \\ C_{n1} & \cdots & C_{nn} \end{bmatrix}$$

In which,

$$\begin{bmatrix} M_{11} & \cdots & M_{1n} \\ \vdots & \ddots & \vdots \\ M_{n1} & \cdots & M_{nn} \end{bmatrix} =$$

$$\begin{bmatrix} M_{11}^1 & \cdots & M_{1n}^1 \\ \vdots & \ddots & \vdots \\ M_{n1}^1 & \cdots & M_{nn}^1 \end{bmatrix} \ldots + \cdots \begin{bmatrix} M_{11}^n & \cdots & M_{1n}^n \\ \vdots & \ddots & \vdots \\ M_{n1}^n & \cdots & M_{nn}^n \end{bmatrix},$$

The general money matrix can be decomposed into a series of specific money matrix.

In general, through the representation of the matrix, it is more intuitive to understand the essence of money as information. Essentially, money is similar to the initial and final information state of goods and services, and its expression in mathematics is consistent with the information state of goods and services. Money is the medium and bridge for the human society to share goods and services, is essentially information. Therefore, when analyzing any money form and future money development trend, we may need to understand the money from the perspective of information in order to better grasp the nature of money.

**4 Money maybe originated from centralized physical ledger**

It is generally believed that money originated from such physical money as shells, livestock, etc. However, recent monetary history studies have shown that the money maybe originated from physical ledger. It is generally believed that after the transition from direct exchange of goods to the money stage, human society has entered the commodity money stage represented by livestock, shells, sugar, salt, and metals, and then entered the stage of credit money represented by banknotes and electronic money. However, recent human archaeology found that before the commodity money, money may exist as an ancient centralized physical ledger (Maurer et al (2013)). Archaeologists discovered a large number of cuneiform records recorded in clay mud when excavating the ruins of the Sumerian temple in the Babylonian region. Analyzed by human archeologists, these cuneiform clay plates are not used for other purposes, but are probably the oldest ledger used for people's commodity trading. These ancient ledgers may have been used as money by humans before commodity money such as shells, salt, and metal coin. That is to say, the centralized physical ledger represented by the clay mud recorded in Sumerian cuneiform writing may be the earliest form of money (Hudson et al (2000)).

The ancient centralized physical ledger was earlier than the commodity money. Similar discoveries have been made elsewhere. Felix Martin, a member of the think tank of the European Stability Program and a Ph.D. in economics at the University of Oxford in England, mentions in his book "Money: The unauthorised biography" that residents of Yap in Micronesia use giant stone wheels as money (Martin (2013)). Because the giant stone wheels are bulky, it is not easy to move. The trading parties cannot hold the wheels, and they cannot be used as commodity money like shells and metal. The giant stone wheels may only act as a ledger, record the island's wealth information. The stone wheels are understood to be a more accurate form of the physical ledger. The great economist John Maynard Keynes, inspired by the stone wheels currency, has revolutionized the world's monetary and financial outlooks. He believes Yap's practice is more logical and the modern approach of gold reserves can be derived from Learn a lot of experience (Keynes (1950)).

Physical ledger as money earlier than the commodity money has a certain logical rationale. Physical ledger may be the fairest, easiest and cheapest way for ancient people to share their labor outcomes. Imagine ancient human society, when the scale of the society is not large enough and people's activities are small, for example, when the population is less than Dunbar's number –somewhere between 150-300 (Dunbar's number is a suggested cognitive limit to the number of people with whom one can maintain stable social relationships), the ancient Sumerian cuneiform ledger or Yap stone wheels ledger to meet the needs of the exchange and sharing of labor produces. People only need to manage centralized physical ledger through certain mechanisms to ensure that the ledger updates can meet the needs of commodity trading activities, and that the information on the ledger is not maliciously tampered with. However, commodity money requires people to work to produce certain commodities that are widely accepted as money. These commodity currencies as store of value are not consumed by people and are almost "idle". To some extent, human labor that produces commodity money is "wasted". In general, since physical ledger don't need to sacrifice a certain amount of special commodities as money, thus saving people's labor, and in the process of commodity exchange, centralized physical ledgers are also very convenient. This may be the most favorable and lowest cost method for society at that time. From the perspective of human self-interest, human society may initially choose a centralized physical ledger as money rather than a special commodity as money.

**5 Physical money is distributed physical ledger**

Physical money is the money that can be seen and touched, and can be perceived by human sense, such as commodity money, gold, silver, banknotes, coins, and so on.

The ancient physical ledger has evolved into physical money, so as to meet the needs of human society for more frequent commodity trading. Physical money can be considered as a huge ledger with a large enough space to cover all users. As the number of human social groups continues to increase and the range of activities continues to expand, ancient centralized physical ledger can no longer meet people's trading needs. For this reason, the centralized physical ledger space is disintegrated and dispersed into a distributed physical money ledger to solve the problem that the performance of the centralized physical ledger cannot meet the needs. This distributed processing method has many similarities with the distributed database and distributed ledger processing methods in the current computer technology field, and solves the problem of insufficient or even safe performance of the central database processing method. Therefore, it can be said that physical money is still the ledger money, which is generalized ledger money. It can be imagined that the space size of distributed ledger is so large that it can cover all users, and it is a giant physical ledger that is recorded in many fixed denominations. When the commodity money, gold and silver are not owned by anyone, and they are not part of the ledger, they cannot be called money. Only the existence of an ownership relationship like the form of ledger can be considered as money. In general, the physical money appears to exist in the form of a single individual, but it can be considered as a physical ledger with a particularly large spatial size, which may be a better explanation for understanding money as information rather than as universal equivalent commodity. This interpretation also unifies the various forms of currency and finds its inherent commonality, that is, the money can be uniformly considered as a ledger. The physical money is also the ledger money.

In Hart Keith ' book " The Memory Bank: Money in an Unequal World," this book states that physical money, as a collective memory bank, records various experiences in human interaction (Hart (2000)). Recently, Maurer Bill wrote in the article "Money as Token and Money as Record in Distributed Accounts" that the physical money is no different from the digital transaction records of contemporary electronic payment methods, like credit cards, PayPal, or even Like Bitcoin, all are recording transactions and human activities (Maurer (2017)).

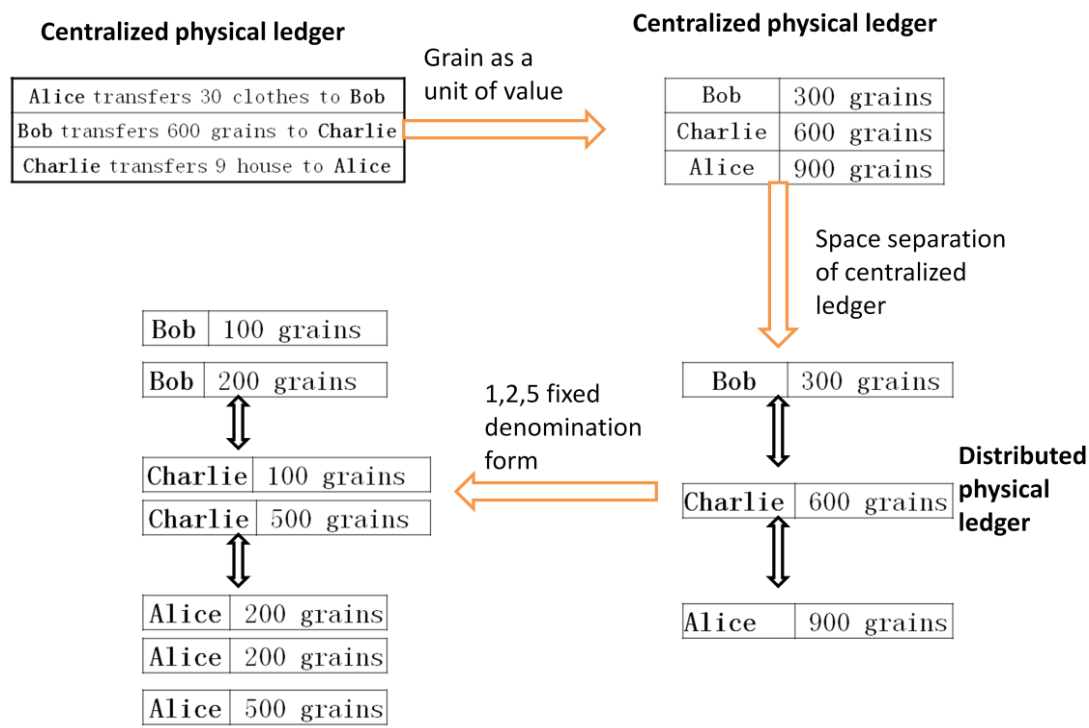

Figure 1: Physical ledger evolves into physical money diagrams. Physical money is the spatial separation of centralized physical ledger, and physical objects are only a carrier for recording ledger, which can be shells, sugar, metal, paper, chip, etc. Grain is used here as the unit of value, and are finally expressed by fixed denomination of 1, 2, and 5.

Since money is a kind of ledger, and physical money is also the ledger, then dividing money into commodity and credit money by value may need to be reconsidered. The value of money itself is not a key point. The key point is on money as ledger information and cannot be maliciously tampered with. This argument that Commodity money has intrinsic value and credit money has not intrinsic value, may not be an objective fact and does not reflect the nature of money. In essence, because both commodity money and credit money can be understood as ledger, the most important thing is not whether there is value, but cannot be maliciously tampered with and forged. It can be inferred that in the commodity money era, there is no suitable intangible product with no intrinsic value or less intrinsic value can be used as the physical money. In fact, in the credit money period, money also has intrinsic value. For example, a beautiful design pattern on a banknote has a high aesthetic value and cultural value. In the past, it was held that the intrinsic value of shells as a commodity money was because shells could be used as ornaments and symbols of good luck. However, if from the view of this kind of value of money, the value of banknotes will not be inferior to shells. Therefore, it is not necessarily true that credit money has no intrinsic value.

From the perspective of anti-counterfeiting technology, we can further deepen our understanding of the nature of money. Recently, in the research of banknote anti-counterfeiting technology, Organic Light-Emitting Diode (OLED) technology has always been a research focus, which can greatly increase the level of existing first-line anti-counterfeiting technology and make it easier for the public to identify (Samsung (2010)), (Andreas et al (2006)). Researcher recently studied inorganic luminescence technology for banknote anti-counterfeiting (Ronald et al (2017)). It is conceivable that if these anti-counterfeiting display technologies can be applied to banknotes, the banknotes can also be used as displays, and can also have their intrinsic value. Therefore, distinguishing money from value may not grasp the nature of money. The nature of money maybe is information, and the prevention of malicious tampering with information may be the core. This also explains the reason why credit money and commodity money alternate in the history of money development, as well as long-term coexistence. What kind of money people choose may not depend on whether it has value or not, but because the ledger information cannot be maliciously tampered with. Evolving from commodity currency to credit currency, the traditional understanding of the history of money development may need to be reconsidered.

**6 Electronic money is the electronic form of physical ledger**

Electronic money is the ledger information stored in the electronic computer etc. Through communication technology, users transmit electronic identity authentication information to access the ledger. Under the protection of the security mechanism, the ledger information is not maliciously tampered with, and its information updates truly reflect the economic activities. The electronic money system is mainly composed of three parts: the front-end customer end, the intermediate information transmission channel, and the back-end ledger system. According to the different front-end customers, they can be divided into two-dimensional code payment, bank card payment, and biometric identification payment etc. According to different intermediate information transmission channels, they can be divided into Internet payment, special network payment, mobile payment, etc. The ledger storage method can be divided into two types: centralized and decentralized (more precisely, multi-center). Recently, decentralized or multi-centralized ledger money is generally called a "digital currency". Bitcoin is one of them and it can be considered as a multi-centered ledger.

E-money solves the problem that people cannot personally access centralized physical ledgers and that centralized physical ledgers cannot be quickly updated based on actual commodity transactions due to frequent visits. As mentioned above, when the population of social groups is less than Dunbar's number and the scope of activities is relatively small, people can more easily access centralized physical ledgers, thereby updating physical ledger information and sharing labor fruits based on real transactions. However, when the population of social groups increases and the scope of activities increases, people's access to physical ledger becomes inconvenient, and it becomes difficult to update centralized physical ledger at a high frequency. At this point, centralized physical ledger can no longer meet the needs of people's commodity transactions. There are two possible ways to resolve the limitations of centralized physical ledger. One way is to dismantle the centralized physical ledger and become large enough in space to cover all users. This is physical money; the other way is to increase the accessibility and processing capacity of centralized physical ledger through electronic technology. This is electronic money. The use of electronic technology to store and process centralized physical money can be quickly updated based on frequent economic transaction activities. Moreover, people do not need to visit the ledger personally but access the ledger through the communication technology so as to solve the limitation of the space distance. The electronic money has resulted from the increase in the speed of computer processing and the rapid development of communications technology. With the aid of information technology, people have broken through the limitations of Dunbar's number, and do not need to physically visit physical ledger at any time. They only need to communicate their own identity information through communication technology, and can update the ledger information according to the actual transaction activities.

Since money is essentially a ledger record, only a string of encrypted data strings cannot represent money. The data string can only be used as money if it corresponds to the account book information and the account book cannot be maliciously tampered with. For the electronic money system, no matter how complex the digital coding, how powerful the encryption technology is, if there is no hardware and software protection, digital information can be easily copied. The uniqueness of electronic money needs to rely on ledger that is not maliciously tampered with. Whether it is the traditional centralized ledger system used by financial institutions, including the account systems used by central banks, commercial banks and third-party payments, or electronic cash based on blind signatures, and offline electronic payment methods, these payment methods are in fact dependent on the uniqueness of the ledger (including the centralized ledger and the spatially separated distributed ledger). Blind-signed electronic cash only achieves the purpose of privacy through a special scrambled data string and cannot be independent from the ledger system with unique characteristics (Chaum (1983)). Without the support of unique ledger, blindly signed electronic cash cannot be used as money. Digital information is easy to copy and does not have uniqueness. Earlier electronic money was not encrypted, and the plaintext was used. This was because the hardware facilities that supported electronic money at that time had high barriers to entry and the attackers could not obtain the plaintext, let alone maliciously change the plaintext, just as the paper currency protects the plaintext of the banknotes through anti-counterfeit technology (Steven (2010)). As long as the ledger is not maliciously tampered with, it is not critical whether it is in plaintext or ciphertext. Based on the above knowledge, only randomness information cannot be used as money. A string of data cannot be used as money. This is also mentioned in the study of quantum money. MIT Quantum Monetary Scholar Scott Aaronson has stated that any digital information can be perfectly reproduced. This is also a major headache for software manufacturers and a reason why digital cash cannot exist (Scott, (2012)).

Off-line electronic money is often similar to cash, sometimes called electronic cash. It is a combination of electronic money and physical currency. E-cash is a piece of data stored in a secure chip. It can be regarded as physical money in essence. It is a kind of physical moeny with no public security features, only machine-readable anti-counterfeiting features, and variable denomination information. When studying the machine-readable anti-counterfeiting technology of banknotes, researchers had considered RFID anti-counterfeiting technology. The European Central Bank has worked with Hitachi to study the anti-counterfeiting application of RFID chips embedded into banknotes (Yoshida (2001)). Although due to the limitations of RFID technology, RFID banknotes have not been successfully developed, but we can be inspired. Imagine that if the RFID chip can meet the needs of the banknote application, an RFID chip is embedded in all banknotes, then the RFID chip can not only be used as security for banknotes, but also can be stored in a piece of electronic cash data information, so the banknote can also be as an e-cash, e-cash can also extend the functionality of a banknote. From this perspective, there is no essential difference between cash and e-cash, except that the former records fixed denomination information through printing, while the latter records variable denomination information through microelectronics storage.

An electronic money system consists of three parts: the front-end client end, the intermediate information transmission channel, and the back-end ledger system. The innovation of the payment system also mainly proceeds from these three parts. The client is responsible for the collection of identity authentication information, the entry of transaction data, and the display of related interaction processes and results. By client presenting the identity information to the back-end ledger system, the ledger information modified by the back-end corresponds exactly to the actual individuals and entities. The client supports biometrics such as portraits, fingerprints, and sounds, traditional account-based passwords, and authentication methods based on USBKEY, OTP, SMS verification codes, and 3DS. The core goal of client innovation is to deliver true identity authentication information and help the back-end ledger system to more accurately, quickly and securely achieve the correspondence between the digital world and the physical world, and realizes the update of the money ledger information to reflect the real world without being maliciously tampered with. The innovation of the information transmission channel is from difficult-maintenance and high-cost special-line channel to the current public Internet channel. SSL/TLS and complex SET communication protocols are used to ensure that the identity authentication information is securely transmitted to the back-end bookkeeping system. At the same time, the update ledger information fed back from the ledger system is securely transmitted to the client. As for the innovation of the back-end ledger system, the most notable part at present is the blockchain, more precisely the distributed ledger technology.

**7 Digital currency is a kind of electronic money with multiple participants recording ledger together**

Digital currency is the electronic money based on blockchain technology. Digital currency is also essentially ledger with multiple participants recording (Thomas et al (2015)). Digital currency is the innovation of ledger in the electronic money system. Digital currency is a subset of electronic money (Pieters (2016)).

The distributed ledger can be regarded as a distributed database in nature, but the biggest difference between this kind of database and the classic database is to solve the intentional modification and destruction of the ledger records (Vukoli´c (2017)). Classical databases deal with objectively unavoidable hardware failures such as server crashes or communication failures with other servers, and distributed ledger databases deal with subjective "faults." Because in the real world, people deliberately tamper with accounting information from time to time, the traditional database technology cannot solve this problem, there are certain security risks. As a result, people are very much looking forward to building an accounting system that prevents vandalism. This system is represented by Bitcoin as a digital currency system based on blockchain technology. It is also a distributed ledger system. Therefore, if you consider the application of distributed ledgers in the payment sector, then the core is still an account book, but the bookkeeping method has been innovative in preventing vandalism. The transaction records in Bitcoin cannot be unique if they are separated from the distributed ledger and cannot be used as currency.

The difference between Bitcoin and traditional electronic money is not the contents of the ledger but the participants who record ledger. Bitcoin is based on the distributed ledger technology, which means that a number of participants jointly record ledger; while traditional electronic money is a centralized ledger technology, in which one participant keeps a separate ledger. In 2013, when interviewed the winner of the 2013 Nobel Prize in Economics, Eugene Fama, He said "I have a cursory knowledge of it. To me it seems like I really don't know the difference between bitcoin and a checking account" (Eugene (2013)). In general, digital currency is the bookkeeping innovation of electronic money, and is a subset of electronic money.

**8 Discussions**

Nature of money is a kind of ledger information. Based on the ledger information perspective, the three basic functions (a store of value, a medium of exchange, and a unit of account) of physical money and electronic money can be more profoundly understood. First of all, for the physical money, the store of value function means the ledger information is accurate, long-term preservation, and cannot be maliciously tampered with; the medium of exchange function means, when a transaction occurs in economic activities, the user indicates the identity information through personally holding physical money, and updates the ledger information by exchanging physical money; the unit of account function means, people reach a unified value measurement unit through consensus, which can be the weight of grain or the weight of gold. It can even set a value unit, such as "yuan", for value comparison and statistics. Secondly, for e-money, the store of value function means the ledger system is stored in the electronic computer etc. Through various security technologies and management, the electronic ledger information is accurate, long-term preservation, and cannot be maliciously tampered with; the medium of exchange function means, when a transaction occurs in the economic activity, the user transfers information such as identity authentication and transaction to the ledger system and obtains authorization to update the ledger information according to the actual transaction activity; the unit of account function of the electronic money are the same as the function of the physical money.

Based on the intrinsic information characteristics of physical money and electronic money, their respective advantages and disadvantages maybe are more obvious. The physical money is a ledger that people can perceive to be recorded in a tangible print way (such as banknotes) or mechanically suppressed (such as coins). Electronic money is an electronic ledger that people cannot perceive as invisible electronic storage. The physical money is accessed and updated in a distributed manner, and electronic money is still accessed and updated in a centralized manner. Physical money and electronic money are the products that evolved from two separate technical paths when the traditional centralized physical ledger performance cannot meet people's transaction requirements: one path is along the space separation of the centralized physical ledger, and the other path is the electronic processing to the centralized physical ledger. The two have their own advantages, and they are not a substitute and an upgrade relationship. Physical money and electronic money are not antagonistic, but they coexist and merge with each other.

First, in terms of security, both physical money and electronic money have their own advantages. The physical money requires fewer infrastructures than the electronic money, and requires less time and scenarios. The physical money can be used all-weather, all-terrain, and even bad natural disasters. The electronic money requires the authorization of the owner to change, so it is in the anti-robbery. Second, in terms of performance, electronic money has always been considered superior performance, but cyber security and network blocking will inevitably occur. Occasionally, electronic devices will fail. At this time, physical money is often superior to electronic money and its use is more convenient and faster. Third, in terms of fairness and privacy, physical money is only relevant to both parties to the transaction, and censorship resistance. However electronic money is subject to censorship by third-party agencies. The physical money has the obvious privacy protection advantages over the electronic money. The physical money can meet the needs of people of different cultural levels and different physical conditions better than electronic money. Fourth, in terms of money value, because money should not be classified according to value and credit currency is not worthless, so money can maximize its other value. Whether it is physical money or electronic money, it can be improved in terms of cultural communication, artistic expression, etc. to increase other value functions beyond money function.

**9 Conclusion**

Money was invented to address the difficulty in the double coincidence of wants between the supply and demand when people exchanged their goods and services. People have goods and services that can be regarded as an informational state of human society. In order to pursue a happy and beautiful life, people share the fruits of labor through exchange of goods and services, and this state of information changes to achieve another kind of information state. Money was invented for this change of state of information. This paper analyzes the process of changing the state of information through the representation of the matrix, and intuitively reveals that money is a kind of information and is the essence of a reliable book record. This article also starts from money history and banknote anti-counterfeiting technology, and analyzes the common characteristics of physical money, electronic money, and digital money. That is, money is a reliable ledger. Physical money and electronic money are two possible ways to solve the problem of centralized physical ledger with limited updating ability and remote access distance. One way is to dismantle the centralized physical ledger and make the space size large enough to cover all users. This is the physical money; the other way is to increase the accessibility and processing capacity of centralized ledger through electronic technology. This is electronic money. The physical money is a distributed form of a centralized physical ledger, which can be understood as a physical ledger with a large spatial size. Electronic money is the electronic form of physical ledger. The ledger is stored and processed by computers et al. The client passes information such as identity authentication through communication technologies and obtains authorization to update the ledger. Digital currency is the bookkeeping innovation of electronic money. It is an electronic ledger that is shared by multiple parties and is a subset of electronic money. Money maybe is the ledger information, only account-based, not value-based, token-based, wallet-based etc.